 \documentclass[final,3p,times,twocolumn,sort&compress]{elsarticle}

 \usepackage{graphicx,amssymb,amsthm,amsmath,lineno,natbib}
 \usepackage{dcolumn,multirow}
 \usepackage[center]{subfigure}
 \usepackage{threeparttable}

\biboptions{square}

\journal{Physics Letters A}

\begin{document}

\begin{frontmatter}



\title{A non-absorbing SIR stochastic lattice gas model on hybrid lattices}


\author[ufsa]{Carlos Handrey Araujo Ferraz\corref{cor1}}
\ead{handrey@ufersa.edu.br}
\cortext[cor1]{Corresponding author}
\address[ufsa]{Exact and Natural Sciences Center, Universidade Federal Rural do Semi-\'Arido-UFERSA, PO Box 0137, CEP 59625-900, Mossor\'o, RN, Brazil}

\begin{abstract}

In this paper, we perform Monte Carlo calculations to study the critical behavior of the spread of infectious diseases through a novel approach to the SIR epidemiological model. A stochastic lattice gas version of the model was applied on hybrid lattices which, in turn, are generated from typical square lattices when inserting a connection probability $p$ that a given lattice site has both first- and second-nearest neighbor interactions. By combining percolation theory and finite-size scaling analysis, we estimate both the critical threshold and leading critical exponent ratios of the non-absorbing SIR model in different cases of hybrid lattices. An analysis of the average size of the percolating cluster and the size distribution of non-percolating clusters of recovered individuals was carried out to determine the universality class of the model.

\end{abstract}

\begin{keyword}


non-absorbing SIR model \sep hybrid lattices \sep percolating cluster \sep critical exponent ratios \sep Monte Carlo simulation
\end{keyword}

\end{frontmatter}


\section{Introduction \label{sec:int}}

In the late 1920s, Kermack and McKendrick \cite{Kermack1927} proposed a system of ordinary differential equations for determining the temporal evolution of the population of individuals who interacted with each other when exposed to an infectious disease. These individuals were initially divided into three classes, namely, susceptible (S), infected (I), and recovered (R) individuals. Taking the acronym for the classes, the model became known as the SIR model. Over the years, variations of this model (e.g., SIRS, SIS, SEIR models) have been proposed considering the different classes of individuals that constitute a given population. The relevance of these approximations cannot be underestimated because they have been successful in modeling the most varied forms of epidemics, including cholera \cite{Hartley2005,Pascual2006}, measles\cite{Kassem2010,Tilahun2020}, rubella \cite{Amaku2003,Prawoto2020}, hepatitis \cite{Zou2010,Khan2013}, influenza \cite{Kim2020}, AIDS \cite{Huang2011,Bashir2017}, COVID-19 \cite{Atkeson2020,Yang2021,Gounane2021}, among many others. Epidemiological models encompass both features of collective dynamics \cite{Wang2016,Shao2022} and complex systems \cite{Helbing2015}; hence, they are important tools for obtaining information regarding the rate of disease spread and for testing protocols adopted by public bodies to contain or mitigate such diseases.

The SIR model is one of the simplest epidemiological models, in which the infected individuals can be removed from the dynamics either by permanent immunity or death. This peculiarity makes the model suitable for simulating the epidemic outbreaks of influenza, SARS, AIDS, etc. The SIR model is in the same universality class as that of dynamic percolation \cite{Grassberger1983, Munoz1999, Ziff2010,Souza2011,Pastor2015}. This study considers a stochastic lattice-gas version of the SIR model with asynchronous site updates. Similar models have been applied to other population dynamics \cite{Satulovsky1994,Antal2001}. For both synchronous and asynchronous versions, a phase transition occurs when the model's control parameters are varied. This transition is found to be of second order between two distinct regimes: one, in which the population remains susceptible (inactive or endemic regime), and other, in which the disease spreads throughout the network (active or epidemic regime), where a significant portion of the population becomes infected or eventually recovers (immune or dead). At the transition point, the system becomes critical and corresponds to an epidemic outbreak threshold. It is worth remarking that the connection between the stochastic and deterministic descriptions (via coupled differential equations) can be achieved through mean-field approximation \cite{Lugo2008,Tome2010}, which results in the so-called Langevin equations associated with the Fokker--Planck equation that describes the temporal evolution of the probability density of the system states.

The lattice gas method \cite{Thorne2007} used herein has been widely used to address problems involving fluid dynamics \cite{Frisch1986,Lallemand1987}, general diffusion processes \cite{Kehr1984}, damage spreading \cite{Ferraz2008,Ferraz2018}, and transport phenomena \cite{Deutsch1996}. This method basically consists of discretization both in time and space, where sites in the network are considered as the most likely particle locations. In molecular dynamics, for example, we only have a temporal discretization in the particle state description. The lattice gas method is especially useful when the dynamics of a particle system needs to be described without considering the detailed microscopic aspects of that system. These aspects are often irrelevant to the description of the general behavior of the system. In this study, we will employ this method to describe the spread of the cloud of recovered individuals over time as the epidemic evolves. 

In particular, an analysis of the average size of the percolating cluster and the size distribution of non-percolating clusters of recovered individuals will be performed using the Newman--Ziff algorithm \cite{Newman2001} to determine the critical exponent ratios of the model. The SIR stochastic lattice gas model is an absorbing-like model because its active phase is characterized by an infinite number of absorbing configurations, in which the final system state only comprises recovered individuals. However, in this work, the simulations are halted as long as the existence of the percolating cluster (spanning cluster) in the system is verified, thereby setting the non-absorbing state of the model. This procedure speeds up the analysis of the simulation data without compromising the guarantee of reaching the asymptotic limit of the system. Only a single percolation cluster is generated in the critical regime of the system. By exploiting the analysis of these clusters, we will determine the universality class of the model. The network topology represents an important feature of the system and directly affects the dynamics of the involved processes\cite{Ferraz2008, Ferraz2017, Ferraz2018}; hence, we will study a hybrid network topology in this work.

Hybrid lattices are formed from regular square lattices ($N=L\times L$) by inserting a probability $p$ that a given site has both first- and second-nearest neighbor interactions. Such networks would simulate a more realistic population mainly formed by two types of individuals, i.e., type I with low connectivity and type II with high connectivity. The special cases with $p=0$ and $p=1$ are also treated here and correspond to pure lattices having all nodes with the connectivity of first neighbors and first and second neighbors, respectively. 

This study mainly aims to understand how extended connectivity effects can affect the critical behavior of the non-absorbing SIR model. Since the connection probability $p$ can be understood as a quenched topological disorder, we want to determine if this kind of disorder is relevant to changing the universality class of the model. There are some criteria devised to try to predict if the quenched disorder can change the critical exponents of a given model, such as the Harris criterion \cite{Harris1974} and its refinement, the Harris--Barghathi--Vojta criterion (HBV) \cite{Barghathi2014}. The Harris criterion states that a second-order transition in a $d$-dimensional system, with original correlation length exponent $\nu$, is stable against the quenched spatial disorder if $\nu>2/d$. Whereas the HBV criterion states that quenched topological disorder is irrelevant with respect to the phase transition stability if the system satisfies $\nu>1/a$, where $a$ is the disorder decay exponent that measures how fast coordination number fluctuations decay with increasing system length scale. Nevertheless, such criteria are known to fail \cite{Schrauth2018}.   

The contents of the article are organized as follows. In section \ref{sec:sir}, we outline the SIR stochastic lattice gas model. In section \ref{sec:mcs}, we describe details of the Monte Carlo (MC) simulation background and lattice generation. In section \ref{sec:r}, we present and discuss the results. Finally, in section \ref{sec:c}, we make the conclusions.

\section{SIR stochastic lattice gas model\label{sec:sir}}

\begin{figure*}[!t]
\centering
\begin{minipage}[t]{1.0\linewidth}
\centering
\subfigure[Case $p=0$]{\label{fig:02a}\includegraphics[scale=0.45, angle=0]{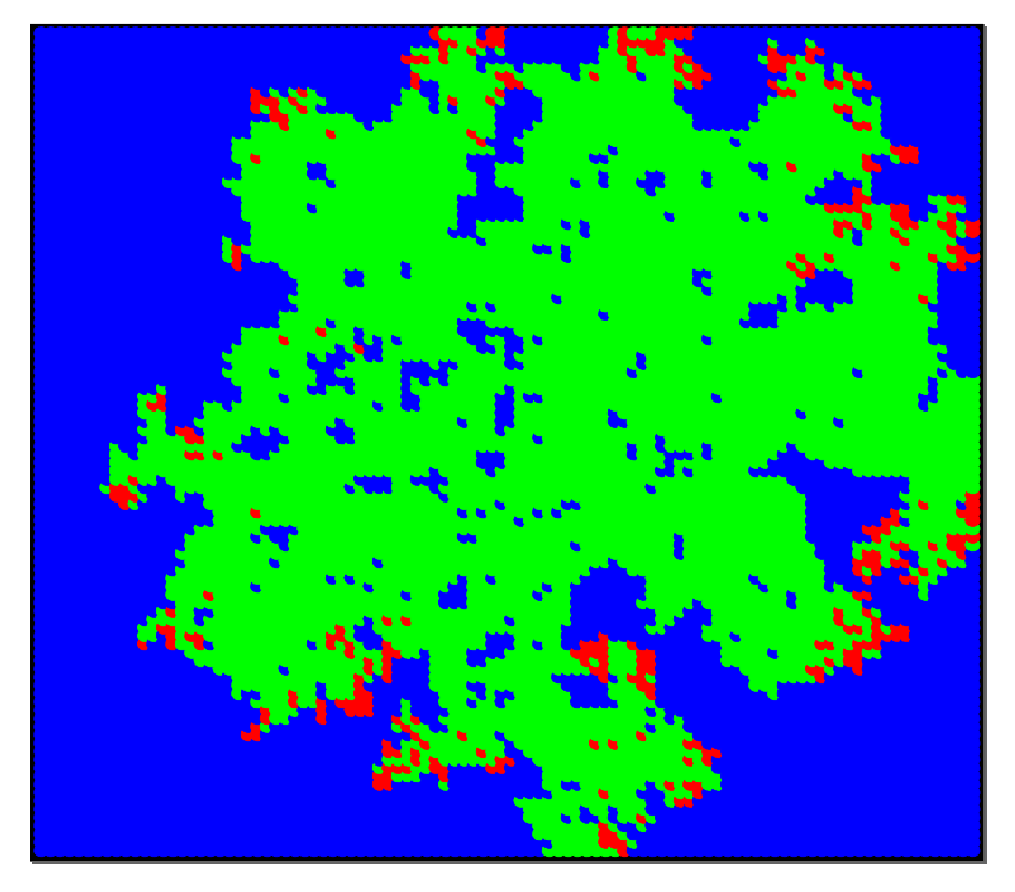}}
\qquad
\subfigure[Case $p=1/2$]{\label{fig:02b}\includegraphics[scale=0.45, angle=0]{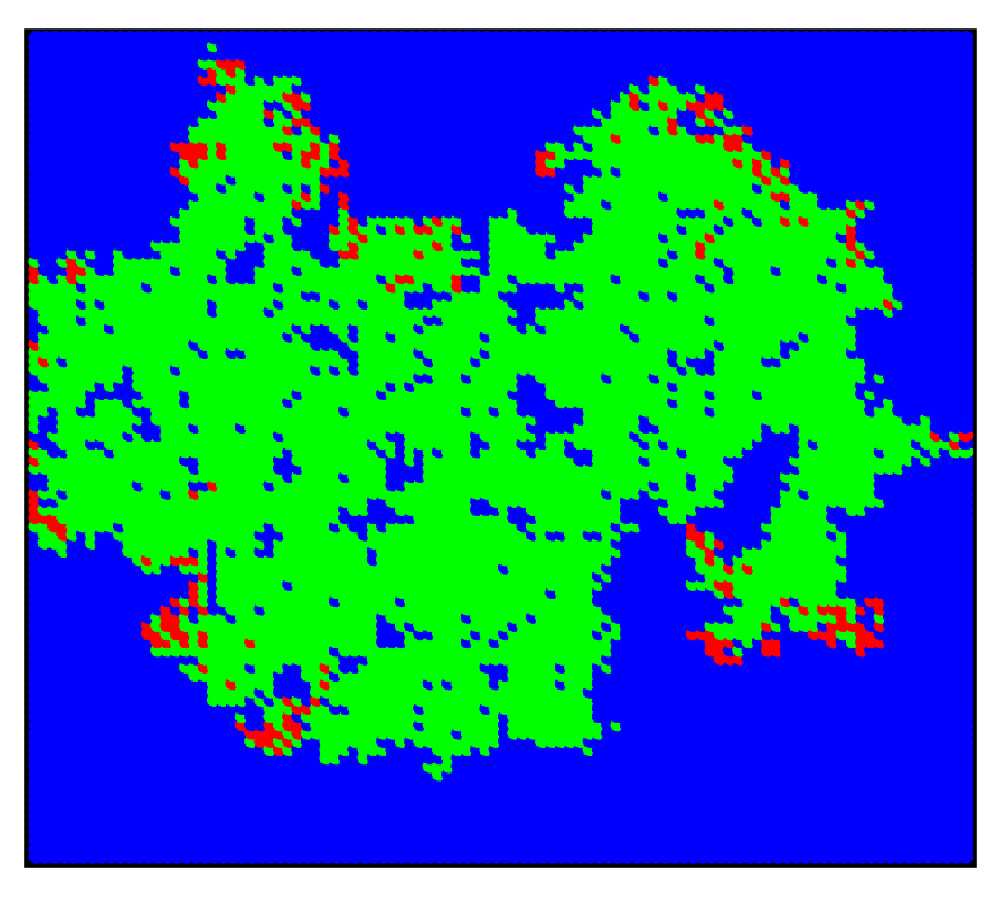}} 
\end{minipage}
\caption{Typical spanning clusters along with few smaller clusters formed close to the epidemic threshold $\lambda _c$ on a $100 \times 100$ lattice for two different $p$ cases. (a) Case $p=0$ (at $\lambda=0.16$) and (b) Case $p=1/2$ (at $\lambda=0.21$).  Susceptible, infected, and recovered individuals are represented by blue, red, and green vertexes, respectively.}\label{fig:02}
\end{figure*}

The SIR stochastic lattice gas model is defined on a given lattice of N sites in which each site can be occupied by just one individual who can be either a susceptible (state $S$), an infected (state $I$), or a recovered (immunized/dead) individual (state $R$). The dynamics consists of two subprocesses, namely, an auto-catalytic one, $I+S\rightarrow I+I$; and a spontaneous one, $I\rightarrow R$. At each time step a site is randomly chosen and then a set of dynamic rules are taken in the following way: \\
i) If the site is in the state $S$ and there is at least one neighboring site in the sate $I$ then the site becomes $I$ with probability proportional to a parameter $\mu$ and the number $z$ of neighboring sites, i.e., $ \mu{\kern 1pt} m /z$, where $m$ is the number of neighboring $I$ sites. \\
ii) If the site is in state $I$ it becomes $R$ spontaneously with probability $\lambda$.\\
iii) If the site is $R$ it remains unchanged.\\
At each site $i$ of a 2D lattice we assign a stochastic variable $\sigma_{i}$ that takes the values 0, 1 or 2, according to whether the site is in the state $S$, $I$, or $R$, respectively. Since the transitions between states in this model are non-equilibrium ones, the allowed transitions of the state $i$ of a site are cyclic, this is, $0 \rightarrow 1 \rightarrow 2$. The corresponding transition rate is represented by $w_{i}(\sigma)$ and describes the transition $\sigma \rightarrow \sigma '$ in which the whole microscopic configuration (microstate) $\sigma ' \equiv (\sigma _1 ,\ldots ,\sigma ' _i ,\ldots ,\sigma _N )$ differs from $\sigma$ only by the state of the $i$-th site. It is given by
\begin{equation} \label{eq:1}
	w_i (\sigma ) = \frac{\mu}
{z}\delta (\sigma _i ,0)\sum\limits_j {\delta (\sigma _j ,1)}  + \lambda{\kern 1pt} \delta (\sigma _i ,1),
\end{equation}
where the summation runs over the nearest neighbors of site $i$ and $\delta(x,y)$ denotes the Kronecker delta. The parameters $\mu$ and $\lambda$ are related to the subprocesses above described, and are chosen such that $\mu+\lambda=1$. 

The system evolves in time according to a master equation for the probability distribution $P(\sigma,t)$ described by
\begin{equation} \label{eq:2}
	\frac{d}{{dt}}P(\sigma ,t) = \sum\limits_i {\{ w_i (\bar \sigma )} P(\bar \sigma ,t) - w_i (\sigma )P(\sigma ,t)\},
\end{equation}
where the microstate $\bar \sigma$ is obtained from $\sigma$ by an anticyclic permutation of the state of the site $i$ ($2 \rightarrow 1 \rightarrow 0$).

\section{ Monte Carlo Simulation and Lattice Generation \label{sec:mcs}}

\begin{table*}[!t] \small
\caption{\label{tab:1} Estimates of the epidemic threshold $\lambda_{c}$ and critical exponents ratios for each $p$ cases.}
\centering
\begin{tabular}{ccccc}
\hline 
 & & & & \\
 Lattice &  Epidemic threshold & $1/\nu$ & $\beta/\nu$ & $\gamma/\nu$  \\ \\ \hline
 & & & &  \\
Square with $p=0$&$\lambda _{c}=0.176(6)$&$0.719\pm 0.011$&$0.109\pm 0.003$&$1.778\pm 0.005$ \\
Hybrid with $p=1/2$&$\lambda _{c}=0.228(4)$&$0.728\pm 0.013$&$0.108\pm 0.003$&$1.790\pm 0.012$ \\
Square with $p=1$&$\lambda _{c}=0.275(0)$&$0.716\pm 0.014$&$0.112\pm 0.003$&$1.776\pm 0.009$ \\
Exact values &$-$&$3/4$&$5/48$&$43/24$ \\
\hline 
 & & & &
\end{tabular}
\end{table*}

We can implement an asynchronous, non-absorbing SIR model on computer by following the kinetic Monte Carlo rules below:
\begin{enumerate}
	\item First, we start with a single central infected site (seed) and the remaining ones being all susceptible on a two-dimensional lattice in which each individual of the population is attached to its respective lattice site. In order to speed up the simulation we create two lists that are updated at each algorithm step: a list of infected individuals (infected list) and a list of recovered individuals (recovered list), which begins empty. 
	\item Next, we update the system state by randomly choosing an available infected site from the infected list and proceed as follows:
	
	\begin{enumerate}
		\item Generate a random number $x$ in the interval $(0,1)$. If $x\leq \lambda$, the infected site is removed from the infected list and placed in the recovered list;
		\item Otherwise (if $x>\lambda$), pick randomly one nearest neighbor of the infected site and make it also infected provided that it is susceptible, adding it to infected list.
	\end{enumerate}
	
	\item Repeat asynchronously the step (2) several times until either there is no infected sites (endemic phase) or there is a percolating cluster of recovered sites (non-absorbing epidemic phase).
\end{enumerate}

One could determine the MC time $t$ by incrementing $t$ by $\delta t=1/n_{I}$, where $n_{I}$ is the current number of infected sites, each time an infected site is pick from the list. However, we do not keep track of time here since we are more interested in static quantities such as the fraction and the mean cluster size of $R$ sites. 

Remarkably, it was shown that the SIR model on square lattices belongs to the same universality class then dynamic percolation (DP). This allows us to investigate the phase transition which takes place in the present non-absorbing SIR model by making use of the percolation theory. Thus we can define the epidemic phase of the model when it is formed a percolating cluster of recovery sites in the system and the endemic phase when it is not. Such a graphical analogy has been used for other compartmental models as well. Following the classical percolation theory it is important first to determine the cluster distribution of recovery sites, i.e., the number of clusters with $s$ recovery sites $n_{p}(s)$. That can be accomplished by using the Newmann--Ziff algorithm, which possess also the built-in feature of identifying whether a percolating cluster was formed or not, depending on the considered $\lambda$ value. Notice that for systems with non-periodic boundaries like the ones concerned here, the percolating cluster is actually a spanning cluster \cite{Newman2001,Sen2011}. 

From the cluster size distribution, we have the fraction of recovery sites in the finite (non-percolating) cluster with $s$ size
\begin{equation} \label{eq:3}
	P_s  = s\frac{{n_p (s)}}{{n_R }},
\end{equation}
 where $n_R$ is the total number of recovery sites and $n_{p}(s)$ is the number of clusters with $s$ recovery sites. Furthermore, the fraction of recovery sites in the percolating cluster $P_\infty$ can be obtained by
\begin{equation} \label{eq:4}
	P_\infty   = 1 - \frac{1}{{n_R }}\sum\limits_s {s{\kern 1pt} n_p (s)},
\end{equation}
such that the above summation excludes the percolation cluster. Now we can define the order parameter from Eq.~(\ref{eq:4}) as 
\begin{equation} \label{eq:5}
	P =  < P_\infty >, 
\end{equation}
where $<x>$ means an average taken over different dynamic realizations. The epidemic phase of the model is reached when $P\neq 0$, that is, when the percolating cluster density is non-zero; while the endemic phase happens when $P=0$. Other important quantities are the mean cluster size
\begin{equation} \label{eq:6}
	S = \frac{1}{{n_R }}\sum\limits_s {s^2 {\kern 1pt} n_p (s)},
\end{equation}
which plays the rule of the susceptibility in classical percolation theory \cite{Kirkpatrick1971,Hoshen1979,Stauffer2014} when taking the average over different runs, i.e.,
\begin{equation} \label{eq:7}
	\chi=<S>,
\end{equation}
the overall mean cluster size
\begin{equation} \label{eq:8}
	S' = \frac{1}{{n_R }}\sideset{}{'}\sum_{s} {s^2 {\kern 1pt} n_p (s)}, 
\end{equation}
and the mean quadratic cluster size
\begin{equation} \label{eq:9}
	M' = \frac{1}{{n_R }}\sideset{}{'}\sum_{s} {s^3 {\kern 1pt} n_p (s)}, 
\end{equation}
where the primed summations above also include the percolating cluster. It is worth remarking that the last two quantities $S'$ and $M'$ only make sense for finite lattice as in the asymptotic limit $(N\rightarrow \infty)$, the percolating cluster size diverges. 

At criticality, the cluster size distribution should obey a power-law scaling \cite{Stauffer1979,Hoshen1979,Souza2011} as
\begin{equation} \label{eq:10}
	n_p (s) = s^{ - \tau } F[s^\alpha\; (\lambda  - \lambda _c )],
\end{equation}
where $\lambda _c$ is the epidemic threshold and $F$ is a scaling function. Similarly, the remaining quantities also obey scaling relations in accordance to classical percolation theory given by
\begin{align}
P &= L^{ - \beta /\nu \;} \tilde P(L^{1/\nu } {\kern 1pt} |\lambda  - \lambda _c |), \label{eq:11}\\
\chi  &= L^{\gamma /\nu \;} \tilde \chi (L^{1/\nu } {\kern 1pt} |\lambda  - \lambda _c |),  \label{eq:12}\\
< S' > & = L^{\gamma /\nu \;} \tilde S'(L^{1/\nu } {\kern 1pt} |\lambda  - \lambda _c |), \label{eq:13}\\
< M' > & = L^{(\beta  + 2\gamma )/\nu \;} \tilde M'(L^{1/\nu } {\kern 1pt} |\lambda  - \lambda _c |). \label{eq:14}
\end{align}
The reciprocal correlation-length exponents $1/\nu$ can be obtained by calculating the modulus of the logarithmic derivative of $P$ at the critical threshold point $\lambda_{C}$
\begin{equation} \label{eq:17}
	\phi \equiv \bigg |\frac{d}{{d\lambda }}\ln (P)\bigg |_{\lambda  = \lambda _c },
\end{equation}
where the derivative of the function $f\equiv \ln(P)$ was evaluated numerically by using a finite central difference scheme in the form
\begin{equation} \label{eq:18}
	\frac{{df}}{{d\lambda }} \simeq \frac{1}{{2h}}\left( {f(\lambda  + h) - f(\lambda  - h)} \right),
\end{equation}
which has an truncation error of the order of $O(h)^2$. The error function $\delta f'$ of $df/d\lambda$ was obtained via error propagation from the uncertainties in the values of $f$ ($\delta f$), being expressed by
\begin{equation} \label{eq:19}
	\delta f'  = \frac{1}{{2h}}\sqrt {(\delta f(\lambda  + h))^2  + (\delta f(\lambda  - h))^2 }.
\end{equation}
In our computations $h$ was taken equal to $2.0 \times 10^{-3}$. Close to $\lambda_{c}$, the quantity $\phi$ obeys a power-law scaling as
\begin{equation} \label{eq:20}
\phi  = L^{1/\nu } \tilde \phi (L^{1/\nu } |\lambda  - \lambda _c |)(1 + bL^{ -\omega } ),
\end{equation}
where $\tilde \phi$ is a scaling function, $b$ is correlation amplitude and $\omega$ is the non-universal correction-to-scaling exponent. We have inserted a correction-to-scaling term in Eq.~(\ref{eq:20}) to improve the fit quality of the data such that the values of $b$ and $\omega$ are chosen in order to minimize the reduced chi-squared ($\chi^{2}$) of the fits. As we will see, however, only for the case $p=1$, the fit quality is slightly improved by this correction term. The optimal values of $b$ and $\omega$ are given in the figure captions for each $p$ case. 

In addition, we can define a universal quantity $U$ in which the scaling dependencies cancel out by combining Eqs.~(\ref{eq:11}), (\ref{eq:13}), and (\ref{eq:14}) in the following way \cite{Souza2011}
\begin{equation} \label{eq:15}
	U = P\frac{{ < M' > }}{{ < S' > ^2 }},
\end{equation}
being analogous to the Binder cumulant for ferromagnetic spin model \cite{Selke2005,Ferraz2015}, and obeying also a scaling relation
\begin{equation} \label{eq:16}
	U = \tilde U\; (L^{1/\nu } {\kern 1pt} |\lambda  - \lambda _c |).
\end{equation}
The crossing point of the $U$ curves for different lattice sizes allow us to estimate the epidemic threshold $\lambda _c$, whereas a finite-size scaling analysis of the observables $P$, $\chi$ and $\phi$ by using Eqs.~(\ref{eq:11}), (\ref{eq:12}) and (\ref{eq:20}) yields the according critical exponent ratios $\beta /\nu$, $\gamma /\nu$ and $1/\nu$. The leading critical exponents $\beta$, $\gamma$ and $\nu$ define the universality class of the system. 

The hybrid lattices used in the present study were constructed starting from regular square lattices with free boundary conditions. First, we begin from a regular square lattice consisting of nodes linked to their four first nearest neighbors by both outgoing and incoming links. Then, with probability $p$, we connect a chosen site also to their second nearest neighbors. After repeating this procedure for every site, a new lattice is constructed with a density $p$ of nodes with both first- and second-nearest neighbor connections. Such networks would mimic a hypothetical population formed basically by two types of individuals: type I with low connectivity, and type II with high connectivity. The special cases with $p=0$ and $p=1$ correspond to pure lattices with all nodes having the connectivity of first neighbors and first and second neighbors, respectively. Figs.~\ref{fig:02a} and \ref{fig:02b} display typical spanning clusters along with few smaller clusters formed close to $\lambda _c$ for the cases with $p=0$ and $p=1/2$, respectively. These clusters arise from the SIR model dynamics. The spanning cluster is formed when the cloud of recovered individuals reaches any two opposing edges of the lattice, in other words, when such a cloud spans the lattice from one side to the other. We grew more than $10^{5}$ spanning clusters for every considered $\lambda$ value to take reliable averages of the above quantities. On average it took about 26 ms to grow a single spanning cluster like those shown in Fig.~\ref{fig:02} on an 3.70 GHz Intel Xeon workstation.  

\section{\label{sec:r} Results and Discussion}

\begin{figure*}[!t]
\centering
\begin{minipage}[t]{1.0\linewidth}
\centering
\subfigure[]{\label{fig:03a}\includegraphics[scale=0.35, angle=0]{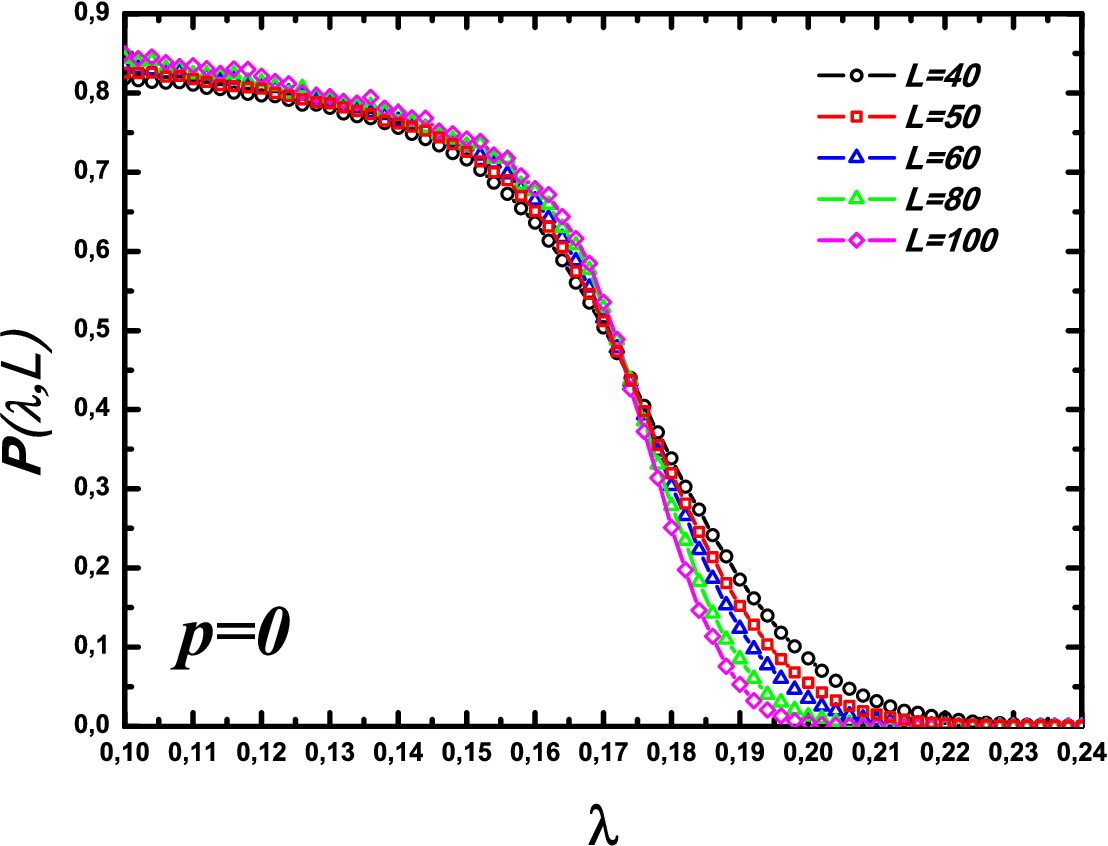}}
\qquad
\subfigure[]{\label{fig:03b}\includegraphics[scale=0.35, angle=0]{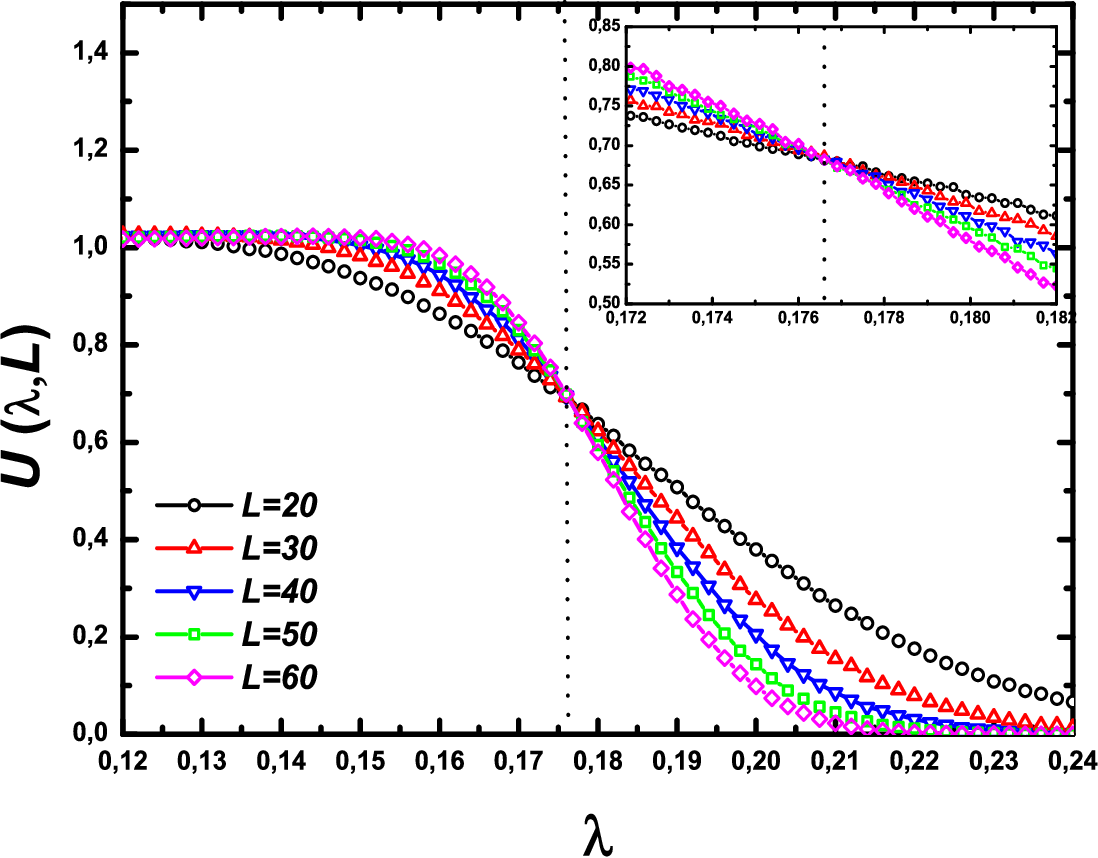}} 
\subfigure[]{\label{fig:03c}\includegraphics[scale=0.35, angle=0]{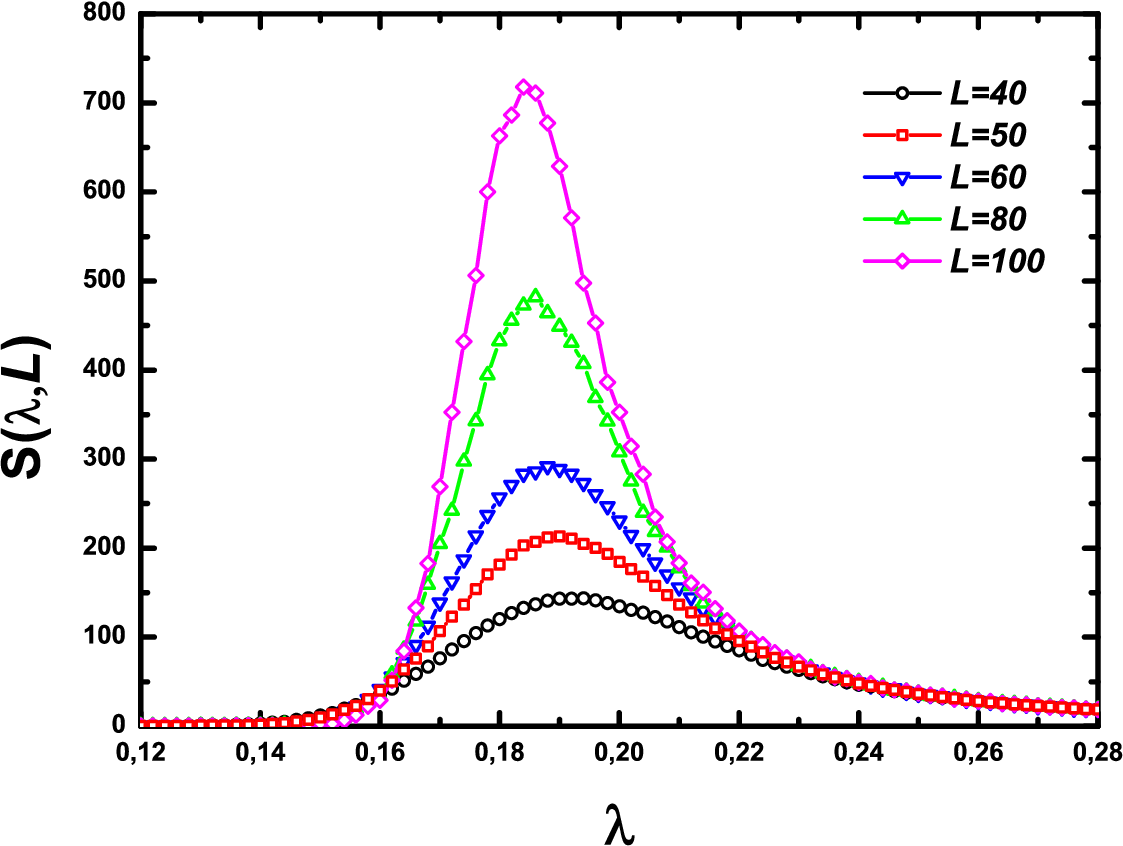}}
\qquad
\subfigure[]{\label{fig:03d}\includegraphics[scale=0.35, angle=0]{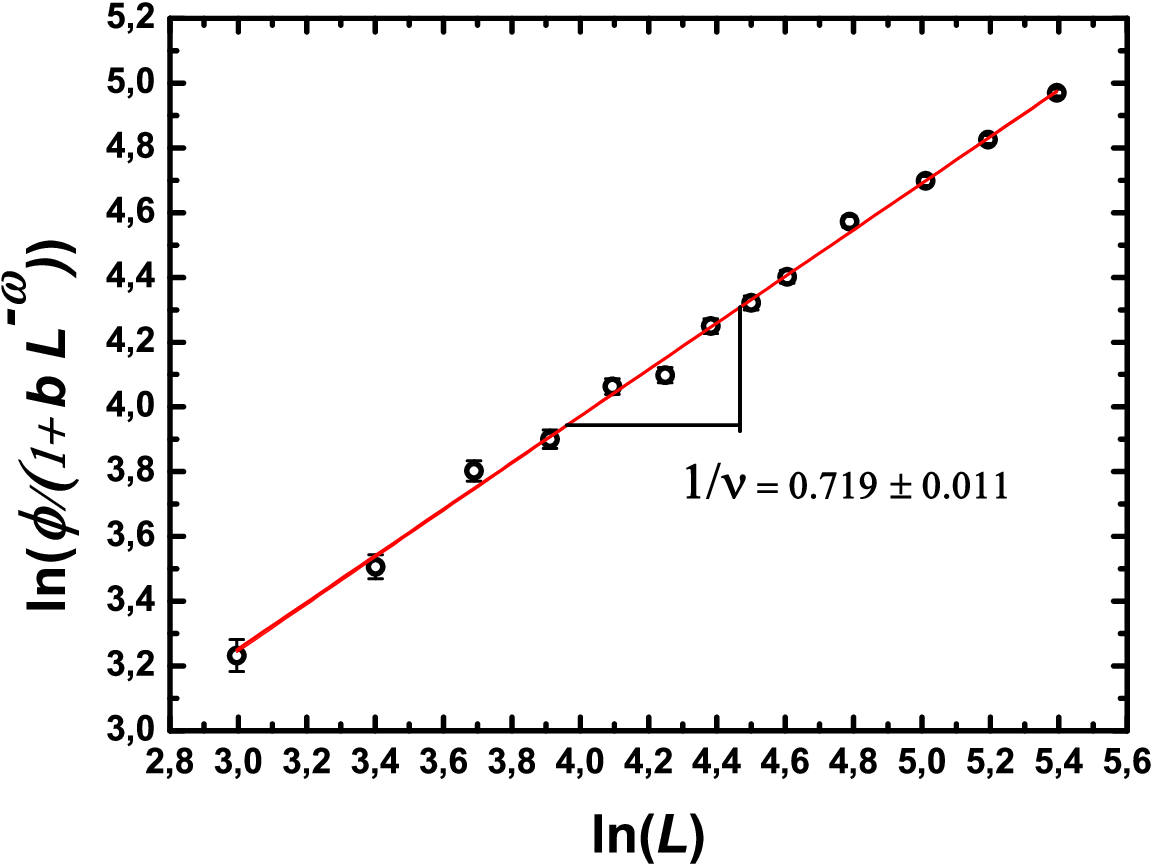}} 
\subfigure[]{\label{fig:03e}\includegraphics[scale=0.35, angle=0]{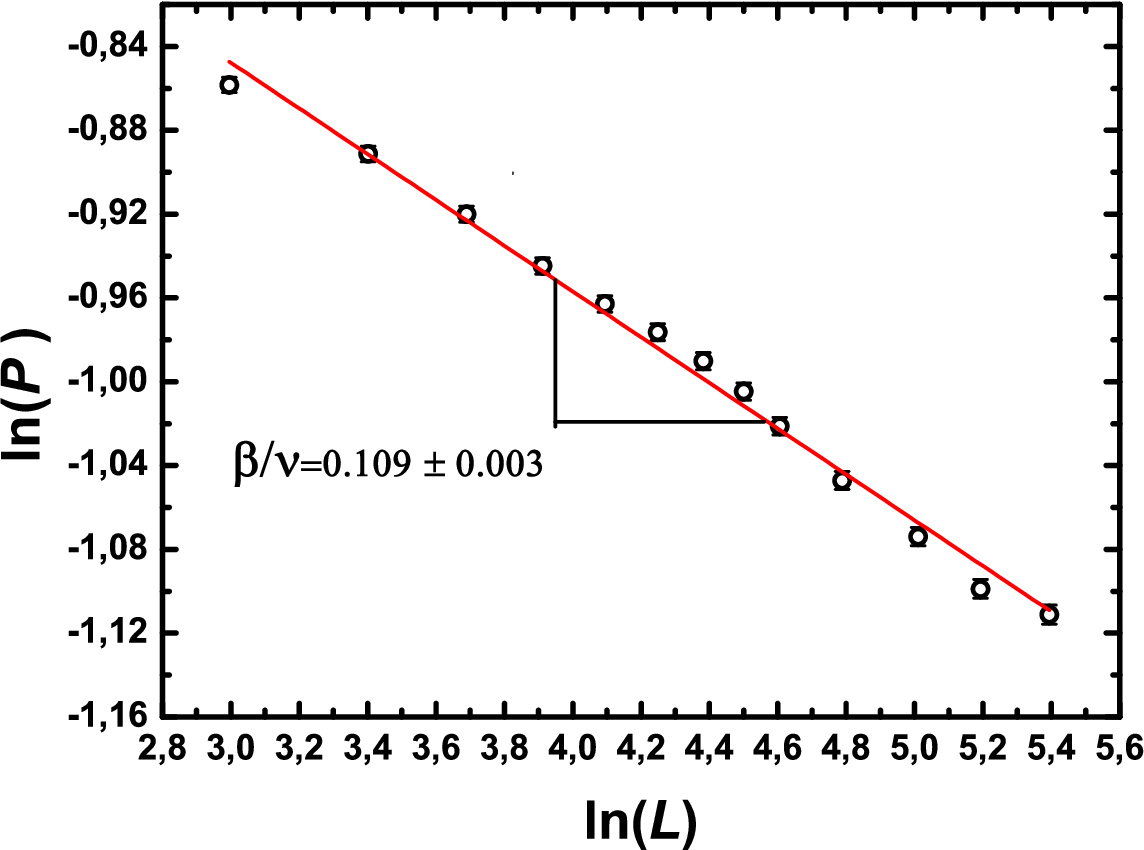}}
\qquad
\subfigure[]{\label{fig:03f}\includegraphics[scale=0.35, angle=0]{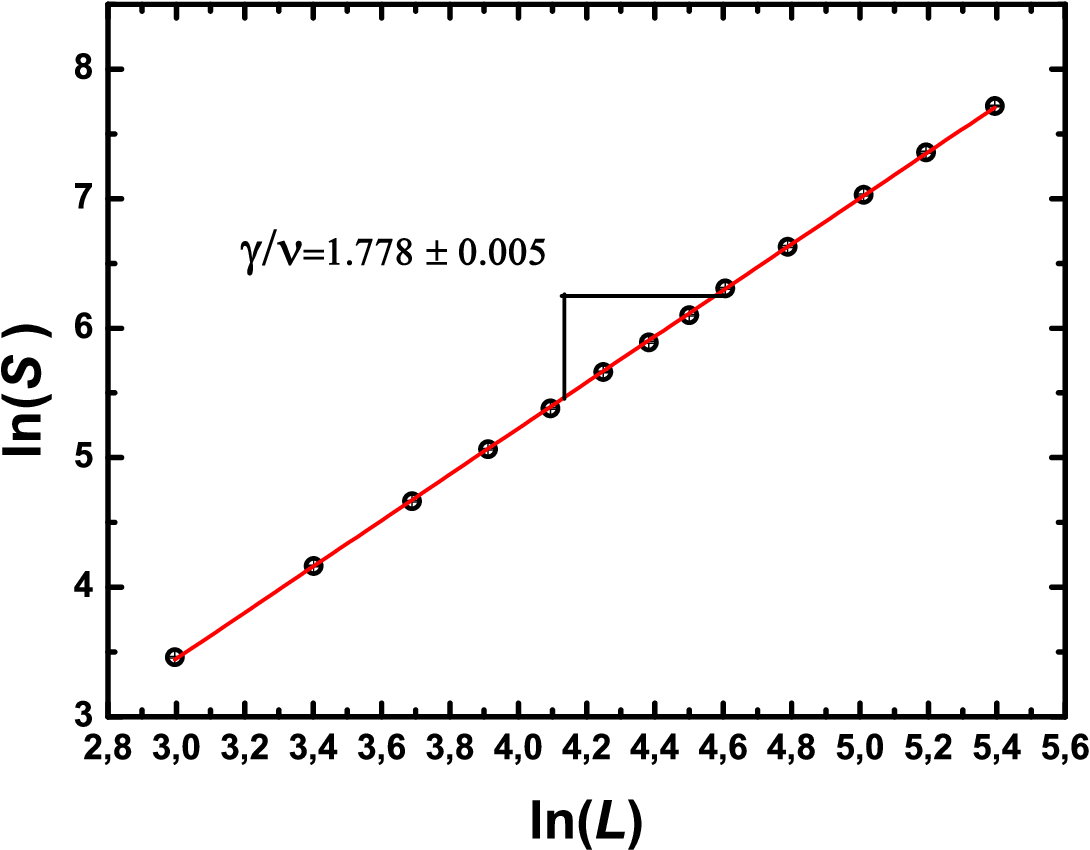}}
\end{minipage}
\caption{Static quantities for the non-absorbing SIR stochastic lattice gas model on square lattices with first-neighbor interactions (case $p=0$). Panels (a), (b), and (c) display the order parameter $P$, Binder cumulant $U$, and susceptibility $\chi$ as a function of the recovery rate $\lambda$, respectively. From the Binder cumulant crossing, we can estimate the epidemic threshold at $\lambda_{c}=0.176(6)$. The inset in panel (b) is a refinement of the calculations for $U$ inside the critical region. Panels (d), (e), and (f) display the log-log plot of the quantities $\phi$ ($b=0$), $P$, and $\chi$ calculated at $\lambda_{c}$ as a function of the linear size of the system $L$, respectively. Red straight lines are the best linear fit to the corresponding data.}\label{fig:03}
\end{figure*}

\begin{figure*}[!t]
\centering
\begin{minipage}[t]{1.0\linewidth}
\centering
\subfigure[]{\label{fig:04a}\includegraphics[scale=0.35, angle=0]{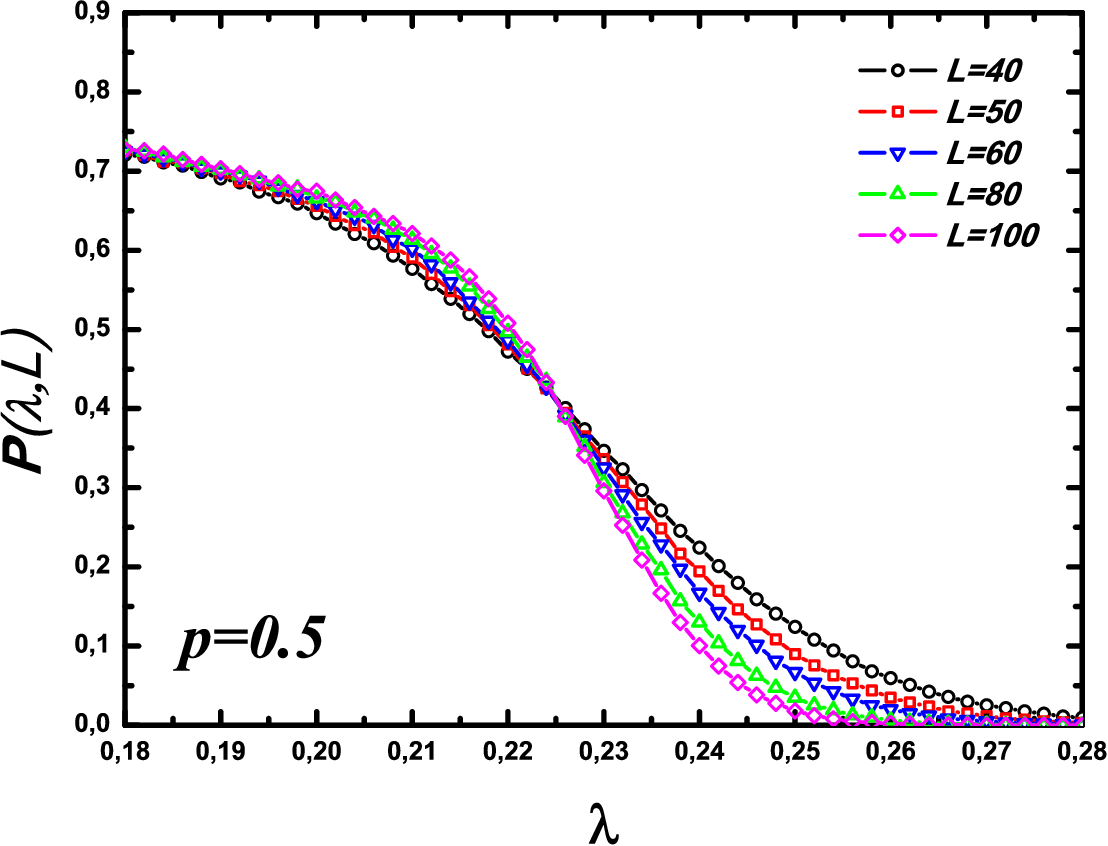}}
\qquad
\subfigure[]{\label{fig:04b}\includegraphics[scale=0.35, angle=0]{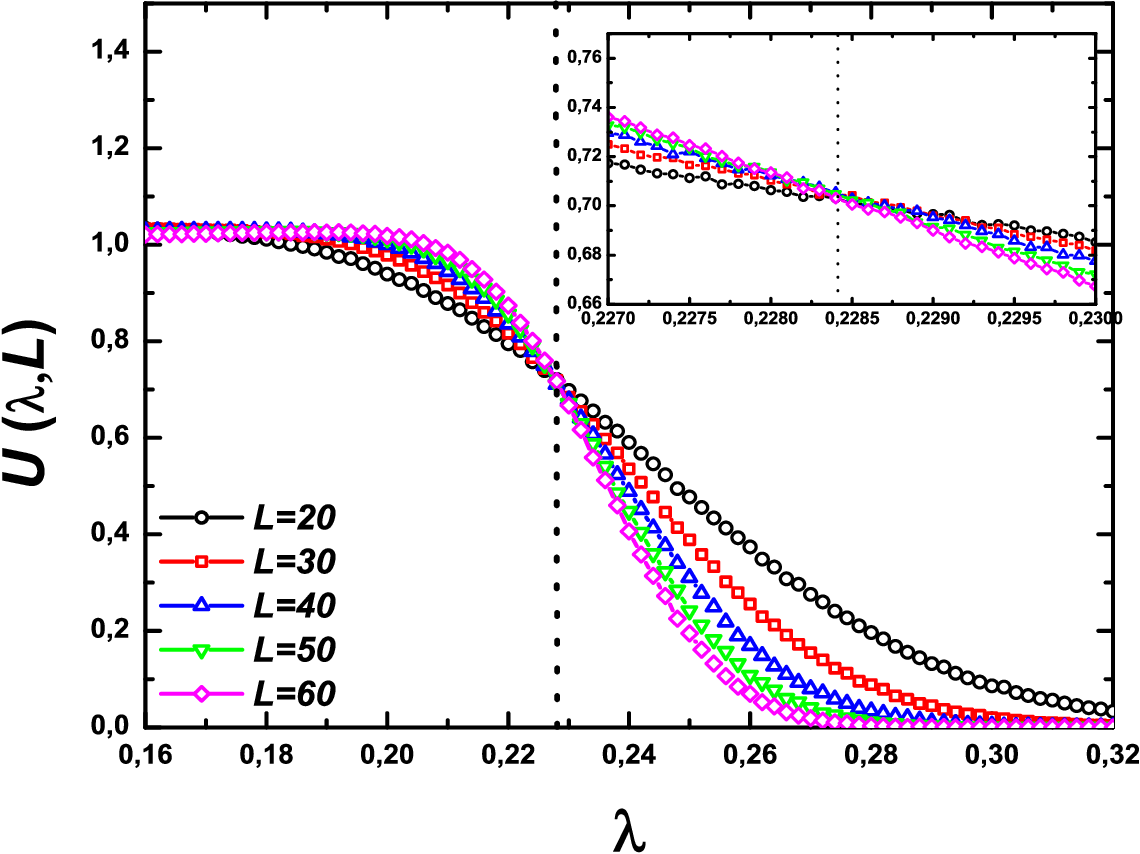}} 
\subfigure[]{\label{fig:04c}\includegraphics[scale=0.35, angle=0]{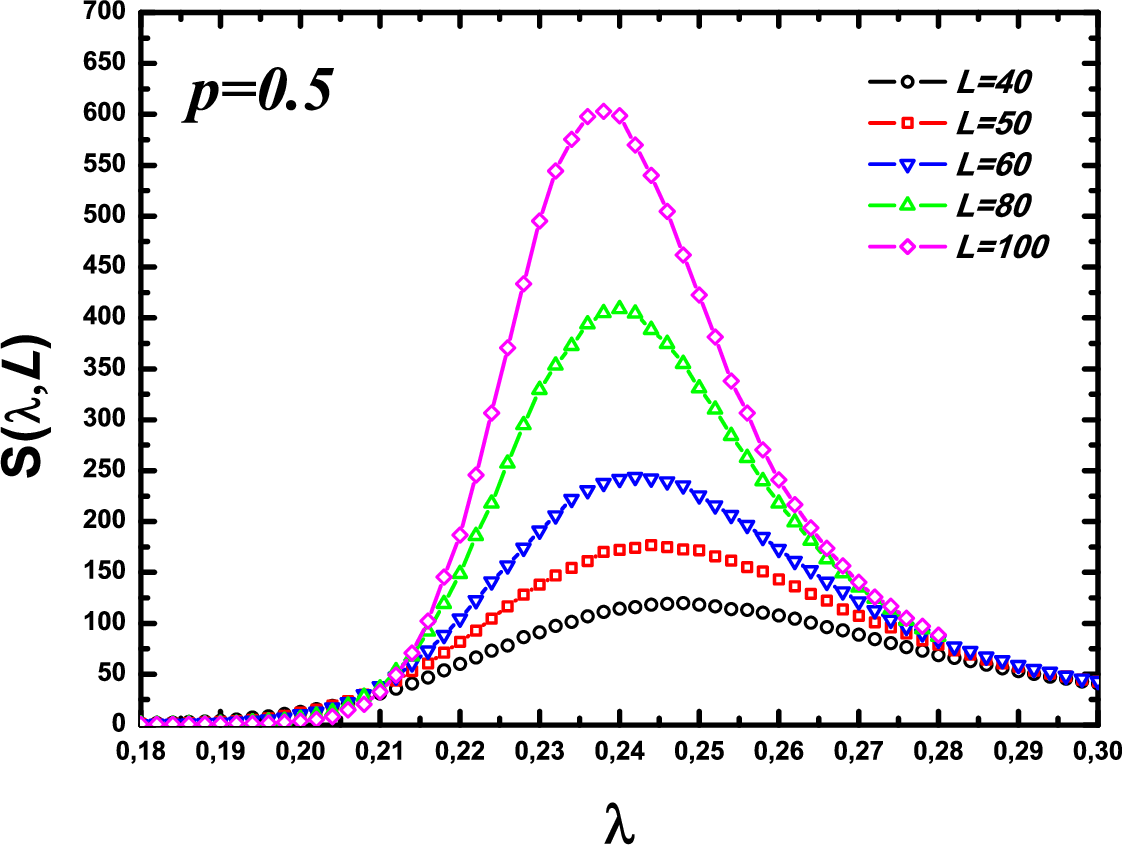}}
\qquad
\subfigure[]{\label{fig:04d}\includegraphics[scale=0.35, angle=0]{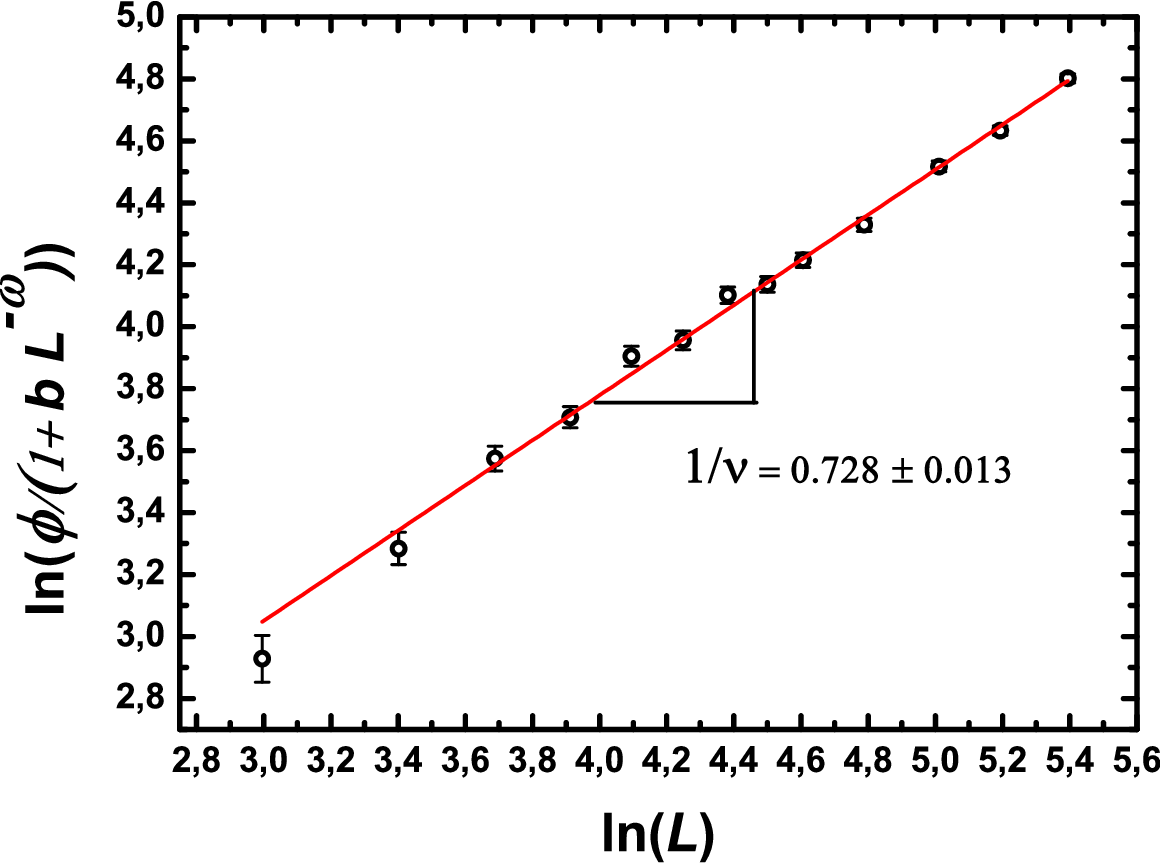}} 
\subfigure[]{\label{fig:04e}\includegraphics[scale=0.35, angle=0]{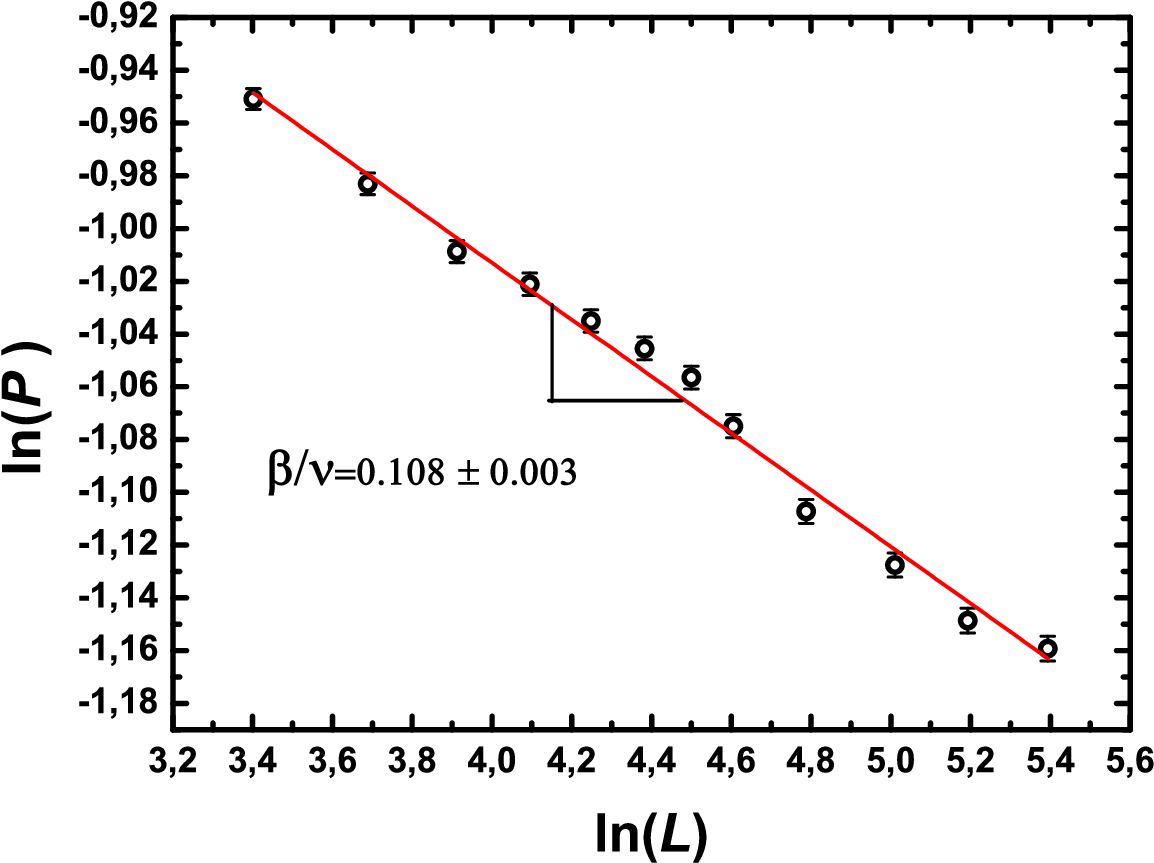}}
\qquad
\subfigure[]{\label{fig:04f}\includegraphics[scale=0.35, angle=0]{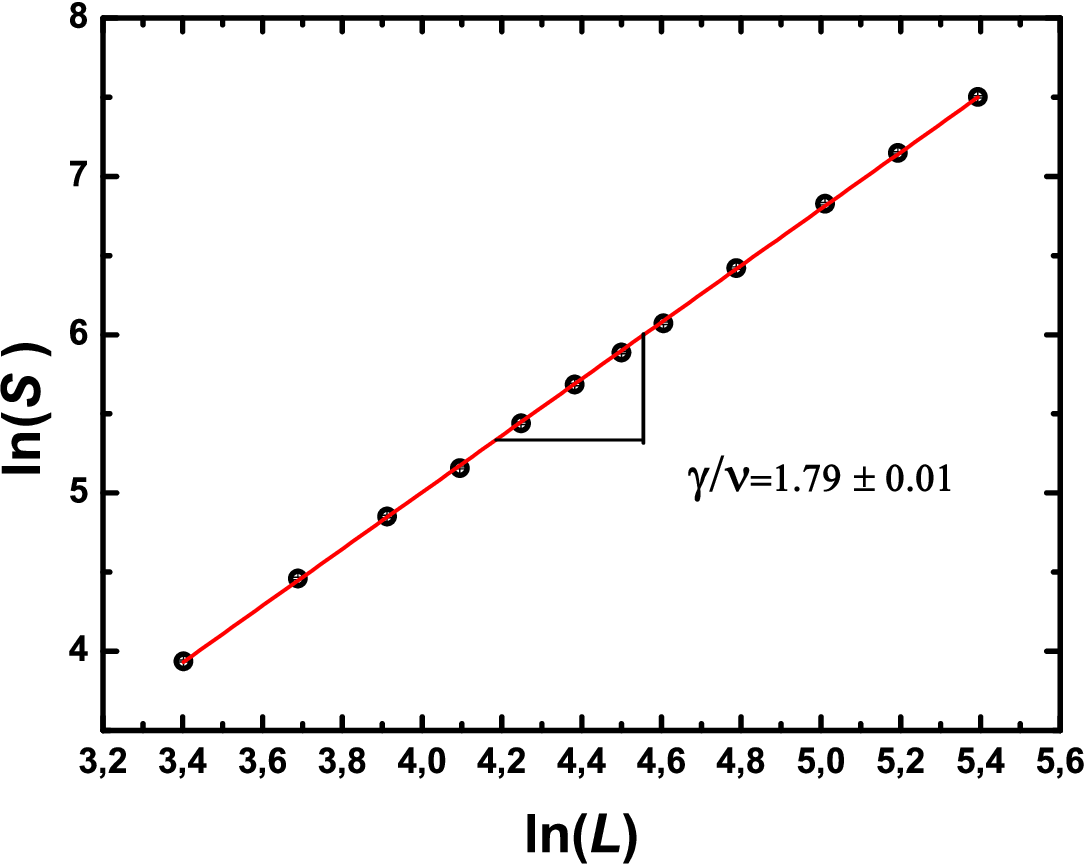}}
\end{minipage}
\caption{Static quantities for the non-absorbing SIR stochastic lattice gas model on hybrid lattices (case $p=1/2$). Panels (a), (b), and (c) display the order parameter $P$, Binder cumulant $U$, and susceptibility $\chi$ as a function of the recovery rate $\lambda$, respectively. From the Binder cumulant crossing, we can estimate the epidemic threshold at $\lambda_{c}=0.228(4)$. The inset in panel (b) is a refinement of the calculations for $U$ inside the critical region. Panels (d), (e), and (f) display the log-log plot of the quantities $\phi$ ($b=0$), $P$, and $\chi$ calculated at $\lambda_{c}$ as a function of the linear size of the system $L$, respectively. Red straight lines are the best linear fit to the corresponding data.}\label{fig:04}
\end{figure*}

\begin{figure*}[!t]
\centering
\begin{minipage}[t]{1.0\linewidth}
\centering
\subfigure[]{\label{fig:05a}\includegraphics[scale=0.35, angle=0]{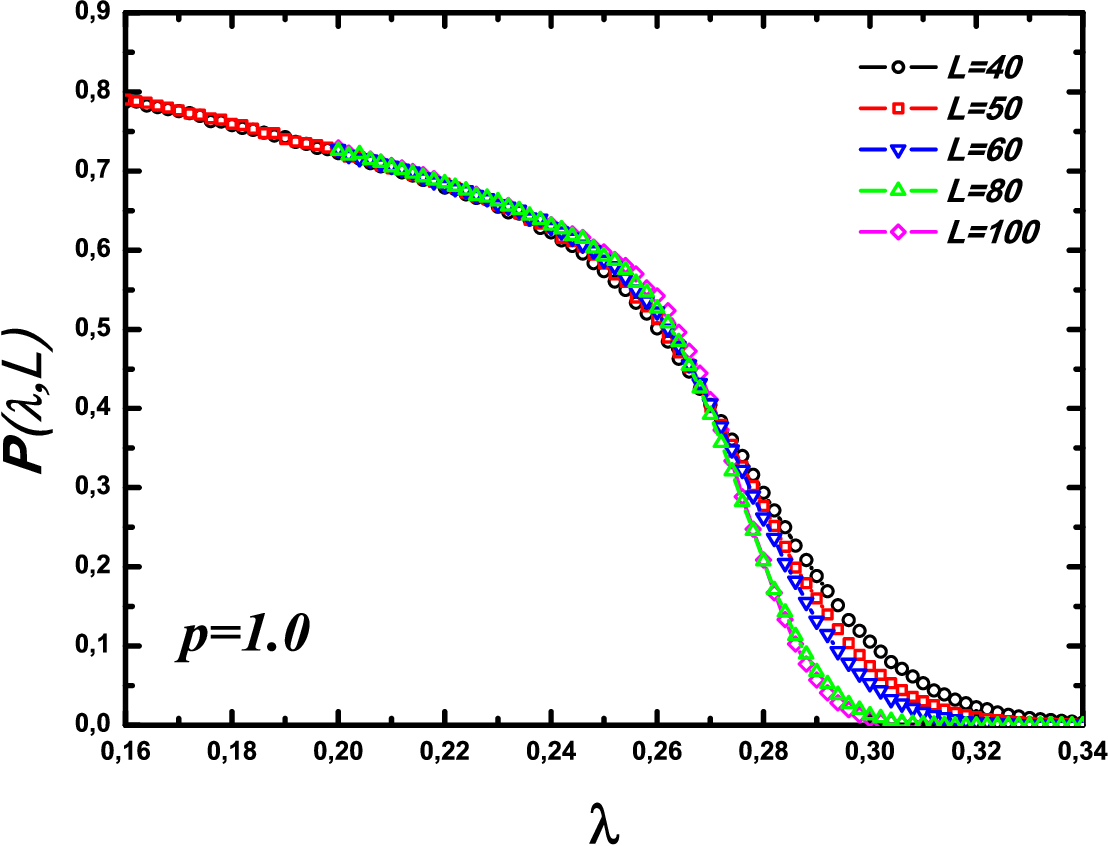}}
\qquad
\subfigure[]{\label{fig:05b}\includegraphics[scale=0.35, angle=0]{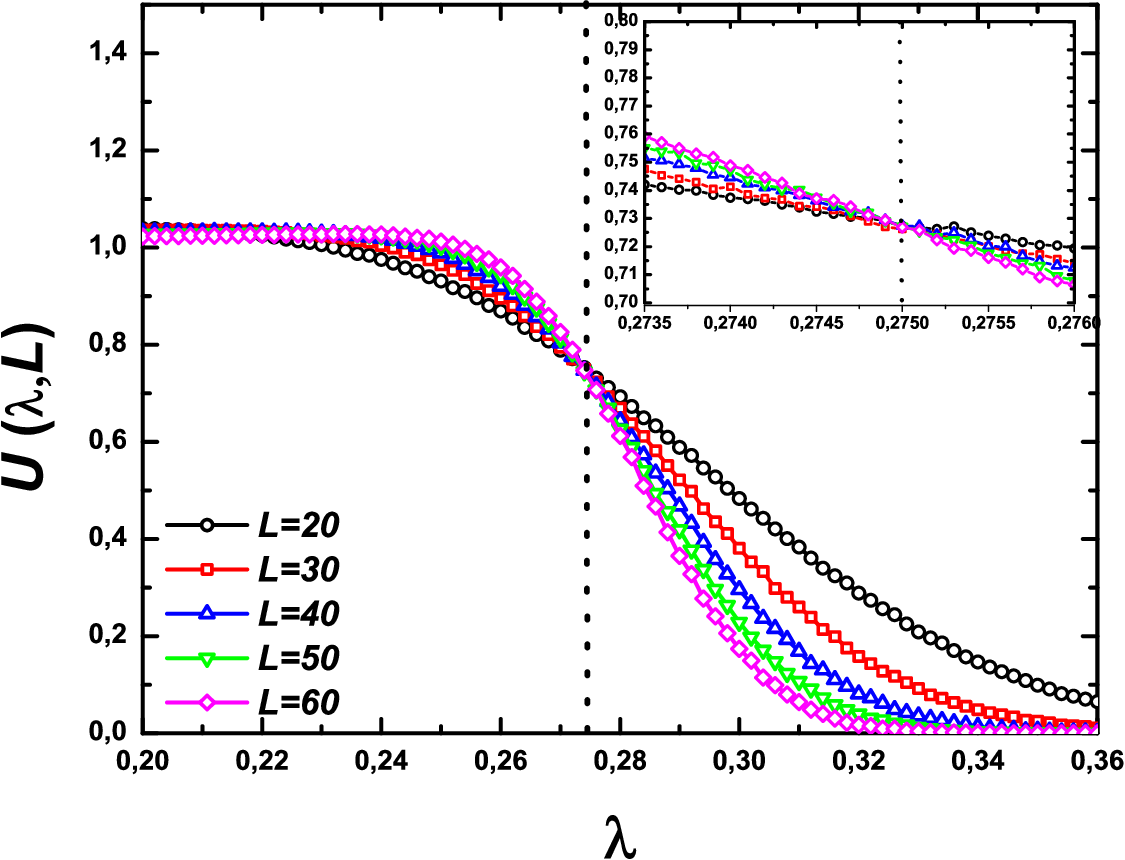}} 
\subfigure[]{\label{fig:05c}\includegraphics[scale=0.35, angle=0]{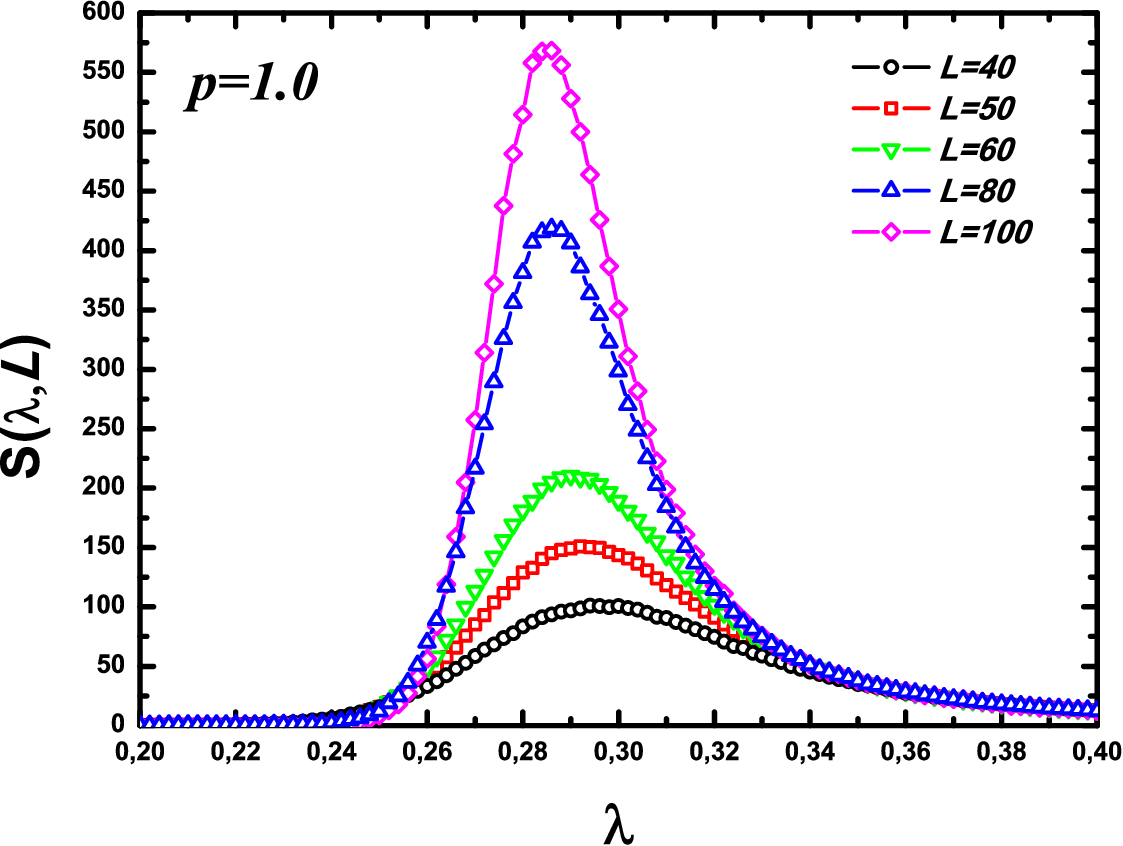}}
\qquad
\subfigure[]{\label{fig:05d}\includegraphics[scale=0.35, angle=0]{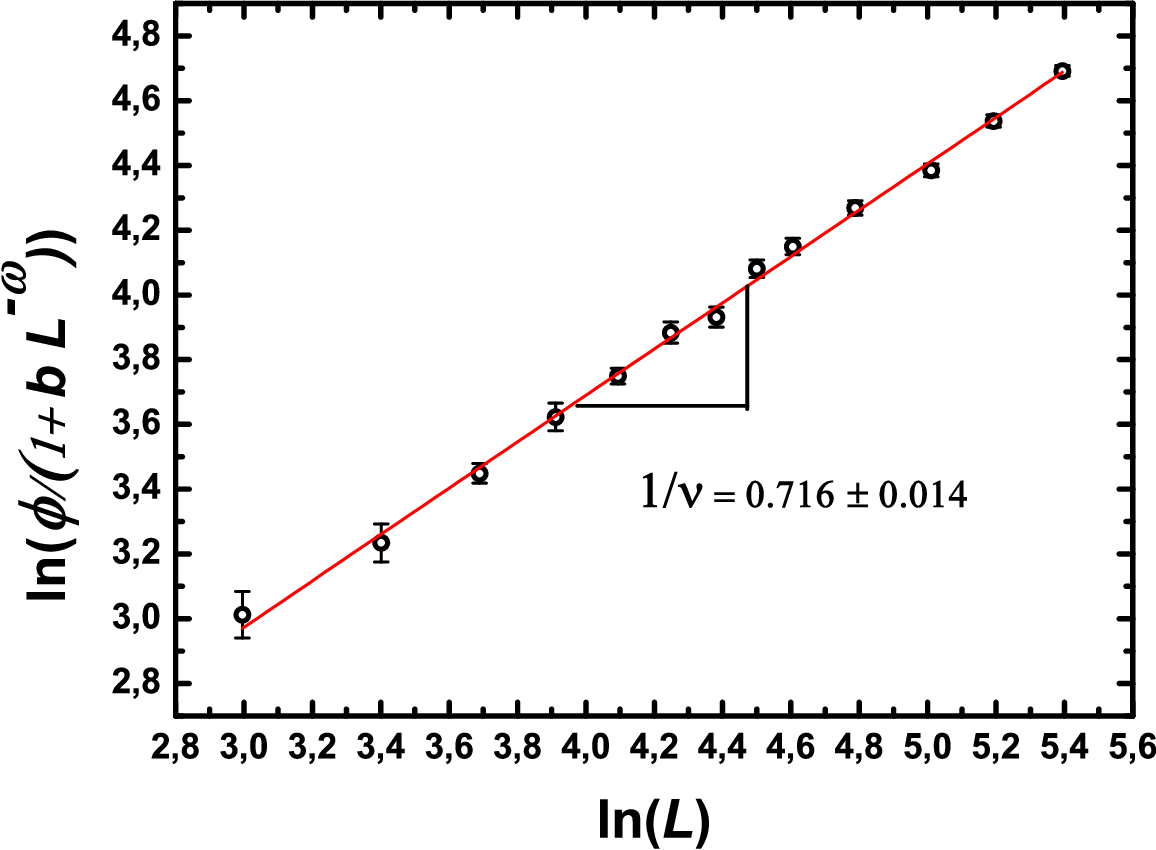}} 
\subfigure[]{\label{fig:05e}\includegraphics[scale=0.35, angle=0]{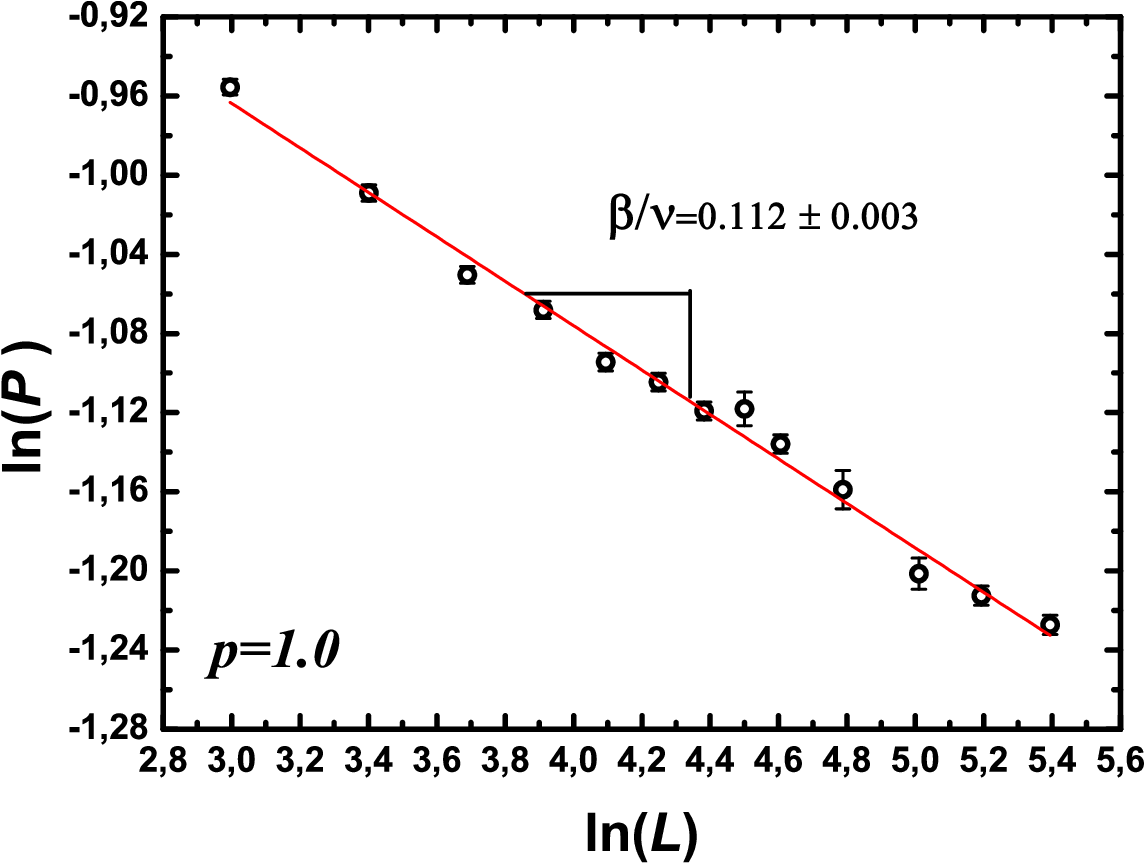}}
\qquad
\subfigure[]{\label{fig:05f}\includegraphics[scale=0.35, angle=0]{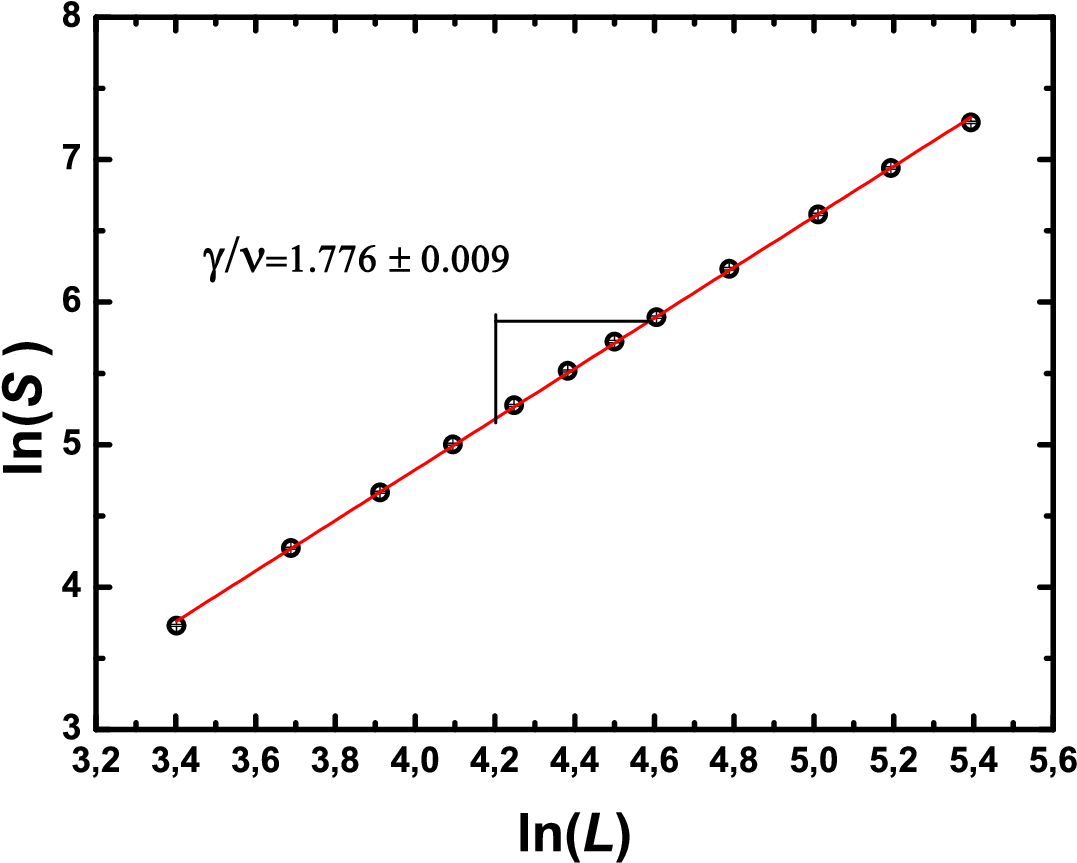}}
\end{minipage}
\caption{Static quantities for the non-absorbing SIR stochastic lattice gas model on square lattices with first- and second-neighbor interactions (case $p=1$). Panels (a), (b), and (c) display the order parameter $P$, Binder cumulant $U$, and susceptibility $\chi$ as a function of the recovery rate $\lambda$, respectively. From the Binder cumulant crossing, we can estimate the epidemic threshold at $\lambda_{c}=0.275(0)$. The inset in panel (b) is a refinement of the calculations for $U$ inside the critical region. Panels (d), (e), and (f) show the log-log plot of the quantities $\phi$ ($b=1.0$, $\omega=1.6$), $P$, and $\chi$ calculated at $\lambda_{c}$ as a function of the linear size of the system $L$, respectively. Red straight lines are the best linear fit to the corresponding data.}\label{fig:05}
\end{figure*}

In this section, we show our numerical results of the non-absorbing SIR model coupled to hybrid lattices. In order to determine both the critical region and the order of the phase transition in this model on hybrid lattices, we calculated the order parameter $P$, Binder cumulant $U$, and susceptibility $\chi$ for each $p$ case in a wide range of the parameter $\lambda$. These quantities were averaged over at least $10^{5}$ different dynamic realizations of the SIR model. Furthermore, for the case $p=1/2$ we consider also different lattice configurations upon taking the averages. We deal with several lattice sizes, ranging from $N=400$ up to $N=48400$. Let us first discuss the case $p=0$. 

\subsection{\label{sec:r1} Case $p=0$}

In Figs.~\ref{fig:03a}, \ref{fig:03b} and \ref{fig:03c} are shown the order parameter, Binder cumulant, and susceptibility as a function of the recovery rate $\lambda$ for the case $p=0$, respectively. As one can see from Fig.~\ref{fig:03a}, a typical second-order phase transition takes place for the case $p=0$. As we shall see, the same conclusion can be drawn for the remaining cases, where a typical sigmoid-shaped curve also occurs. From Binder cumulant crossings, we can estimate the corresponding epidemic thresholds for each case. In the inset of Fig.~\ref{fig:03b} is shown a refinement of the calculations for $U$ inside the critical region. The critical thresholds were estimated with five significant figures. For the case $p=0$, we obtained $\lambda_{c}=0.176(6)$. As expected, it is approximately equal to that observed in the absorbing version of the SIR model. By taking the slope of the log-log plot of the quantity $\phi$ versus the linear size of the system $L$, we can get an estimate for $1/\nu$. Fig.~\ref{fig:03d} shows the best linear fit to Eq.~(\ref{eq:20}) for the case $p=0$. We obtained $1/\nu=0.719\pm0.011$. Error bars were estimated by using Eq.~(\ref{eq:19}). The least reduced $\chi^{2}$ was equal to 1.34, with a goodness-of-fit probability $Q$ of 19,7\%. This value of $1/\nu$ is within few standard deviations from the exact critical exponent ratio $1/\nu=3/4$ of 2D dynamical percolation and it deviates only 4\% from this exact value. Similarly, a finite-size scaling analysis of the magnitudes of the order parameter $P$ and the susceptibility $\chi$ at $\lambda_{c}$ by using Eqs.~(\ref{eq:11}) and (\ref{eq:12}) yield, respectively, the critical exponent ratios $\beta /\nu$ and $\gamma /\nu$. Figs.~\ref{fig:03e} and \ref{fig:03e} show the log-log plot of $P$ and $\chi$ (both calculated at $\lambda_{c}$) against $L$, respectively. The red straight lines in those figures are the best linear fit to Eqs.~(\ref{eq:11}) and (\ref{eq:12}), respectively. We obtained $\beta /\nu=0.109\pm 0.003$ and $\gamma /\nu=1.778\pm0.005$. These values are also in very good agreement with the exact critical exponent ratios of 2D dynamical percolation, namely, $\beta /\nu=5/48$ and $\gamma /\nu=43/24$. These estimates of the critical exponents ratios and critical threshold $\lambda_{c}$ for case $p=0$ are summarized and compared with the corresponding exact values form 2D dynamics percolation in Table \ref{tab:1}.

\subsection{\label{sec:r2} Cases $p\neq 0$}

An analogous analysis can be done for the cases $p=1/2$ and $p=1.0$. Figs.~\ref{fig:04a} and \ref{fig:04a} display the order parameter $P$ as a function of the recovery rate $\lambda$ for the cases $p=1/2$ and $p=1$, respectively. As already remarked, both systems undergo also a second-order transition with their respective $P$ curves exhibiting a typical sigmoidal shape. While Figs.~\ref{fig:04a} and \ref{fig:04a} show the Binder cumulant for $p=1/2$ and $p=1$, respectively. From these figures, one see that the crossing points are located at $\lambda_{c}=0.228(4)$ for $p=1/2$ and $\lambda_{c}=0.275(0)$ for $p=1$. Likewise, we took the slope of the log-log of $\phi$ defined by Eq.~(\ref{eq:17}) versus $L$ to estimate the exponent ratio $1/\nu$ for $p=1/2$ and $p=1$ cases. Figs.~\ref{fig:04d} and \ref{fig:05d} display the linear curve fitting to Eq.~(\ref{eq:20}) for $p=1/2$ and $p=1$ cases, respectively. Error bars were calculated by using Eq.~(\ref{eq:19}). We got $1/\nu=0.728\pm 0.013$ for $p=1/2$ (with a least reduced $\chi^{2}=1.14$ and a goodness-of-fit probability $Q=32.2\%$) and $1/\nu=0.716\pm 0.014$ for $p=1$ (with a least reduced $\chi^{2}=0.68$ and a goodness-of-fit probability $Q=75.5\%$). These estimates deviate, respectively, only 3\% and 4.5\% from the exact value of $1/\nu$. Such results strongly suggest that both cases are in the same universality class of 2D dynamic percolation. Moreover, these results are fairly close to each other, as one falls within only one standard deviation from the other. Similarly, we took the slopes of the log-log plots of $P$ and $\chi$ versus $L$ for both cases. Again, the red straight lines are linear regressions to the corresponding data. From Figs.~\ref{fig:04e} and \ref{fig:05e}, we obtained $\beta /\nu=0.108 \pm 0.003$ for $p=1/2$ and $\beta /\nu=0.112 \pm 0.003$ for $p=1$, respectively. While from Figs.~\ref{fig:04f} and \ref{fig:05f}, we got $\gamma /\nu=1.79\pm 0.01$ for $p=1/2$ and $\gamma /\nu=1.776 \pm 0.009$ for $p=1$, respectively. 
As can be seen, these estimates are rather close to each other. These results clearly suggest that the non-absorbing SIR model on both hybrid lattices ($p=1/2$) and square lattices with first- and second-nearest neighbor interactions (case $p=1$) belong also to the same universality class as that of 2D dynamic percolation. Nevertheless, the critical threshold in the non-absorbing SIR model increases with increasing $p$-connection disorder. The estimates of the critical exponent ratios and the critical threshold $\lambda_{c}$ for all treated $p$ cases are summarized in Table~\ref{tab:1}.  

\section{\label{sec:c} Conclusions}

We performed Monte Carlo simulations of the non-absorbing SIR stochastic lattice gas model on hybrid lattices to study the critical behavior presented by these systems. Both the critical threshold and leading critical exponent ratios were estimated for different cases. Our numerical analysis has revealed that the quenched $p$-connection disorder is irrelevant to changing the critical exponents of the model, irrespective of the considered $p$ value, strongly suggesting that the present model belongs to the same universality class as that of two-dimensional dynamical percolation. However, it was found that the critical threshold in the non-absorbing SIR model increases with increasing $p$-connection disorder. 

This study has wide applications to many issues, including not only the spread of infectious diseases, but also in general diffusion processes, damage propagation in random networks, and performance optimization in multi-core architectures. Finally, we expect that the results presented in this paper can be helpful to understand how topological disorders can affect the critical properties of other related complex systems. 


\bibliographystyle{model1b-num-names}
%



\end{document}